\documentclass[aps,prl,twocolumn,footinbib,groupedaddress,longbibliography]{revtex4-2}
\usepackage{graphicx}
\usepackage{subfigure}
\usepackage{dcolumn}
\usepackage{bm}
\usepackage{times}
\usepackage{amsmath}
\usepackage{amssymb}
\usepackage{float}
\usepackage{color}
\usepackage{multirow}
\usepackage{url,hyperref}
\usepackage[normalem]{ulem}

\begin{document}
\title{Modulating near-field radiative energy and momentum transfer via rotating Weyl
semimetals}

\author{Huimin Zhu,$^{1,4}$ Gaomin Tang,$^{2,*}$ Lei Zhang,$^{1,4,\dagger}$ and Jun Chen$^{3,4,\ddag}$}
\affiliation{$^1$State Key Laboratory of Quantum Optics Technologies and Devices, Institute
  of Laser Spectroscopy, Shanxi University, Taiyuan 030006, China\\
  $^2$Graduate School of China Academy of Engineering Physics, Beijing 100193, China\\
  $^3$State Key Laboratory of Quantum Optics Technologies and Devices, Institute of
  Theoretical Physics, Shanxi University, Taiyuan 030006, China\\
  $^4$Collaborative Innovation Center of Extreme Optics, Shanxi University, Taiyuan
  030006, China}

\bigskip

\begin{abstract}
We study near-field radiative transfer of energy, angular momentum, and linear momentum between a nanoparticle and a plate
consisting of magnetic Weyl semimetals, and demonstrate that these can be efficiently tuned by a relative angle between the Weyl node separations. This tunability originates from the coupling between the particle-induced rotational Poynting vector and the nonreciprocal surface plasmon polaritons supported by the plate. Remarkably, we uncover a counterintuitive regime in which both energy and angular momentum transfer are maximized when the Weyl node separations are antiparallel rather than parallel. This arises from optimal mode matching between the rotation direction of the particle's circular heat flux and the propagation direction of the surface plasmon polaritons in the antiparallel configuration.
\end{abstract}

\maketitle

\textit{Introduction.} Nonreciprocal thermal radiation has emerged as a rapidly developing frontier in nanoscale radiative thermal transport. Established approaches to break reciprocity include applying magnetic fields to magneto-optical materials~\cite{khandekar2019thermal,hu2015surface,2020absence,GT21}, imposing bias voltages on graphene~\cite{correas2019,Ben2016,morgado2022}, and employing time modulation~\cite{GT24,yu2024time,ghanekar2022violation}. These schemes, however, generally require external fields. Magnetic Weyl semimetals (MWSMs) provide an intrinsic alternative. Owing to their topologically nontrivial band structure and intrinsic time-reversal symmetry breaking~\cite{WSM_radiate1,WSM_radiate3,WSM_radiate4,Konabe2024Anomalous}, the momentum-space separation of Weyl nodes acts as an effective internal magnetic field, giving rise to a large anomalous Hall conductivity~\cite{sonowal2019giant,chen2019optical,bugaiko2020surface}. This intrinsic Hall response supports nonreciprocal surface plasmon polariton (SPP) modes, offering distinctive advantages for controlling thermal radiation~\cite{WSM_radiate5,yu2022near,YU2022123339,Naeimi2025,guo2020radiative,YU2023,GT_WSM,ma2023drift}.

In nonreciprocal systems, thermal fluctuations not only mediate radiative heat exchange but also generate lateral forces and torques through the transfer of linear and angular momentum, respectively~\cite{bimonte2017nonequi,bimonte2017nonequi,lateral_17,lateral_21,gao2021thermal,lateral_23,lateral_21-2,zhu2024current}. For example, torque on a single magneto-optical nanoparticle~\cite{gao2021thermal}, lateral forces in nonreciprocal two-plate systems~\cite{lateral_23}, and lateral forces and torques on dipolar particles near nonreciprocal surfaces~\cite{lateral_21-2,zhu2024current} have been reported. Despite these advances, the fundamental mechanisms governing force and torque generation in MWSM nanoparticle-plate systems remain
unexplored. In this paper, we explore an active control scheme for near-field energy and momentum transfer in a MWSM nanoparticle-plate system, achieved by rotating the Weyl node separation of the plate relative to that of the particle. This relative rotation effectively modulates the coupling strength between the particle's circular modes and the nonreciprocal SPP modes supported by the plate. The results reveal a pathway for manipulating nanoparticles with near-field thermal radiation using the intrinsic nonreciprocity of MWSMs.

\begin{figure}
	\centering
	\includegraphics[width=\columnwidth]{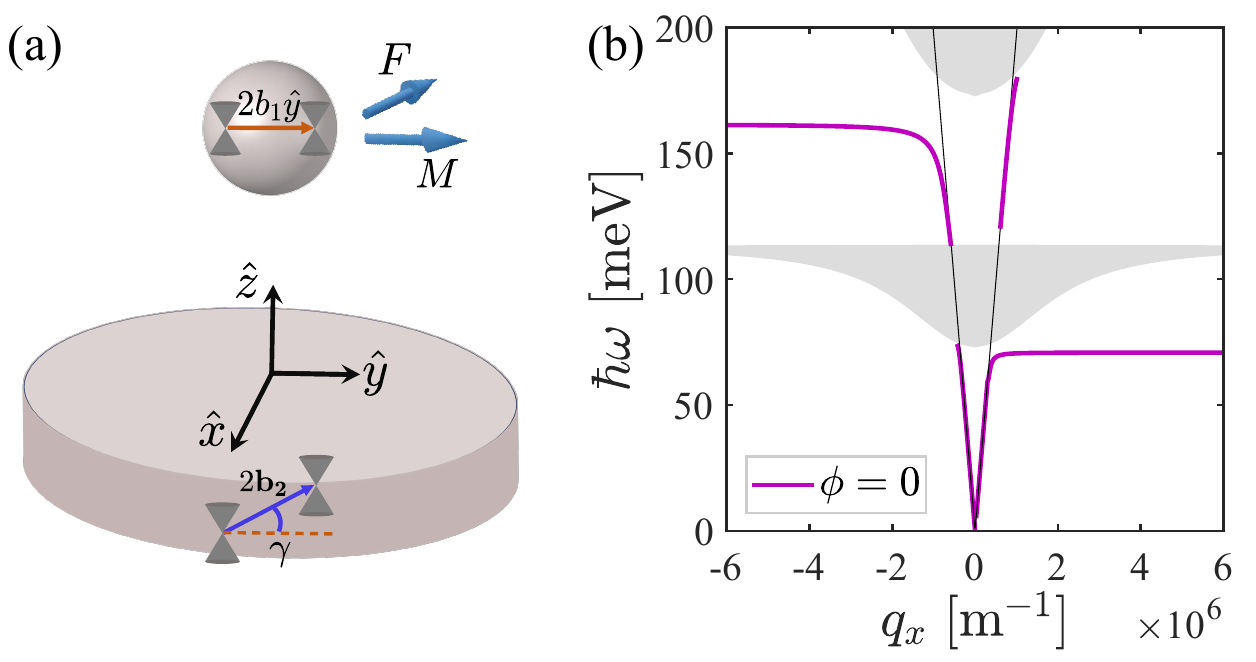} \\
  \caption{(a) Schematic of near-field radiative energy transfer between a MWSM nanoparticle and a MWSM plate, as well as the lateral force $\bm   F$ and torque $\bm  M$  exerted on the nanoparticle. The relative angle $\gamma$ is defined as the angle between the Weyl node separations of the nanoparticle ($2{b}_1 {\hat y}$) and the plate ($2\mathbf{b}_2$), with magnitudes $b_1 = b_2 = b$.
           (b) The SPP dispersion supported by the MWSM plate with the node separation along the $+y$ direction at the azimuthal incidence angle $\phi=0$. The gray regions show the continua of the bulk plasmon modes.}
	\label{fig1}
\end{figure}

\textit{Model and formalism.} The system is illustrated in Fig.~\ref{fig1}(a), where a MWSM nanoparticle of radius $R$ is located at a distance $d$ above the surface of a MWSM plate.
The particle and plate are held at temperatures $T_p$ and $T_e$, respectively. In MWSMs, time-reversal symmetry is broken when a Dirac point splits into a pair of Weyl nodes with opposite chirality. We assume that the momentum separation $2\mathbf{b}$ of the Weyl nodes is along the $+y$ direction and the MWSM possesses no chiral magnetic effect. The corresponding permittivity tensor takes the form~\cite{guo2023light}
\begin{equation}
\label{Eqe}
  \epsilon(\omega) =
  \begin{bmatrix}
    \epsilon_{d} & 0 & i\epsilon_{a} \\
    0 & \epsilon_{d} & 0 \\
    -i\epsilon_{a} & 0 & \epsilon_{d} \\
  \end{bmatrix} ,
\end{equation}
with $\epsilon_{a} = b e^2 / (2\pi^2 \epsilon_{0} \hbar \omega)$ and $\epsilon_{d} = \epsilon_{b} + i\sigma/\omega$. Here, $e$ is the electron charge, $\epsilon_{0}$ is the vacuum permittivity, and $\epsilon_{b}$ and $\sigma$ denote the background permittivity and the bulk conductivity, respectively. The explicit expression for $\sigma$ and the material parameters are provided in the Supplemental Material~\cite{supple}.

When the incidence plane is at an angle $\phi$ with respect to the $x$ axis, the dielectric tensor in the Cartesian coordinate system where the $x'$ axis lies within the incidence plane becomes $\epsilon' = \mathcal{R} \epsilon \mathcal{R}^{T}$ with the rotation matrix
\begin{equation}
  \mathcal{R} =
  \begin{bmatrix}
  \cos\phi & -\sin\phi & 0 \\
   \sin\phi & \cos\phi & 0 \\
   0 & 0 & 1 \\
  \end{bmatrix}.
\end{equation}
Within the framework of fluctuational electrodynamics, the total power $H$, the lateral forces $F_j$ with $j=x,y$ and the lateral torques $M_j$ are given by $P = \int_0^{\infty} {d\omega}\, p(\omega)/{2\pi}$, with $P\in \{H, F_x,F_y, M_x,M_y\}$ and the corresponding spectral densities $p\in \{h,f_x,f_y,m_x,m_y\}$ as~\cite{supple}
\begin{widetext}
\begin{align}
&h(\omega)=4 \hbar\omega k_0^2 \delta n(\omega) \int \frac{d^2 {\bm q}}{(2\pi)^2} \big[ {\rm Im}(\alpha_{xx}) {\rm Im} \left( G_{xx} +G_{zz} \right)+{\rm Im}(\alpha_{yy}) {\rm Im} \left(G_{yy} \right)+ 2{\rm Re}(\alpha_{xz}) {\rm Re}\left(G_{xz} \right) \big], \\
&f_{j}(\omega)= -4\hbar k_0^2 \delta n(\omega) \int \frac{d^2 {\bm q}}{(2\pi)^2} \Big\{q_{j} \big[ {\rm Im}(\alpha_{xx}){\rm Im} \left( G_{xx}+G_{zz} \right) +{\rm Im}(\alpha_{yy}) {\rm Im}\left(G_{yy} \right)+ 2{\rm Re}\left(\alpha_{xz} \right) {\rm Re} \left( G_{xz} \right) \big] \Big\},  \\
&m_x(\omega)=4\hbar k_0^2 \delta n(\omega) \int \frac{d^2 {\bm q}}{(2\pi)^2} \big[ {\rm Im}(\alpha_{xx}+\alpha_{yy}){\rm Re}\left(G_{zy}\right) -{\rm Re}(\alpha_{xz}) {\rm Im}\left( G_{yx} \right) \big],\\
&m_y(\omega)=4\hbar k_0^2 \delta n(\omega) \int \frac{d^2 {\bm q}}{(2\pi)^2}  \big[2 {\rm Im}(\alpha_{xx})  {\rm Re}\left(G_{xz}\right)  + {\rm Re}(\alpha_{xz}) {\rm Im}\left(G_{xx} +G_{zz} \right) \big],
\end{align}
\end{widetext}
where $\bm{q}=(q_x, q_y)$ denotes the in-plane wavevector, and the photon-occupation difference is
\begin{equation}
\delta n(\omega) =[e^{\hbar\omega/k_B T_p}-1]^{-1} - [e^{\hbar\omega/k_B T_e}-1]^{-1}.
\end{equation}
We have abbreviated the Green's function $G_{ij}({\bm q}, d, d, \omega)$ as $G_{ij}$ with $i, j \in\{x, y, z\}$. The Green's function of two points above the plate in Fig.~\ref{fig1}(a) can be written as
\begin{equation}\label{EqG}
G(\mathbf{r}_1,\mathbf{r}_2,\omega)= \int \frac{d^2 {\bm q}}{(2\pi)^2} e^{i{\bm q} \cdot \big(\mathbf{R}_1-\mathbf{R}_2 \big)} G({\bm q},d,d, \omega) ,
\end{equation}
with $\mathbf{r}_1 = (\mathbf{R}_1, d)$, $\mathbf{r}_2 = (\mathbf{R}_2, d)$, and
\begin{align}
G({\bm q},d , d ,\omega)= &\frac{i}{2\beta_0} \left(  \hat{p}_{+} \hat{p}_{+}^T   +  e^{2i \beta_0 d}  r_{p} \hat{p}_{+} \hat{p}_{-}^T  \right) ,
\end{align}
where the polarization vectors are given by
\begin{equation} \label{EqR}
\hat{p}_{\pm}^T=\frac{1}{k_0}\left(
\mp \beta_0 \cos \phi, \
\mp \beta_0 \sin \phi, \
q \right) .
\end{equation}
Here, $r_{p}$ denotes the Fresnel reflection coefficient for $p$ polarization, which dominates the coupling between the particle and the SPP modes of the plate~\cite{Annika2019Radiative}.

Before investigating the near-field radiative transfer, we characterize the optical properties of the MWSM plate and nanoparticle. The dispersion relation of SPPs supported by the plate is given by~\cite{ma2023drift}
\begin{equation}
 \epsilon_{\rm eff }\beta_0+ \beta_1 +i \, \epsilon_a q \, \cos \phi /\epsilon_d = 0,
 \end{equation}
where $\epsilon_{\rm eff}=\epsilon_d - (\epsilon_a \cos \phi)^2/\epsilon_d$, $\beta_0 = \sqrt{k_0^2 -q^2}$, $\beta_1 = \sqrt{\epsilon_{\rm eff} k_0^2 -q^2}$, $q = |\bm{q}|$, and $k_0=\omega/c$.
In the numerical calculation, we use $R = 30\,$nm, $d=0.1\,\mu$m, $T_p=305\,$K, and $T_e=300\,$K. Figure~\ref{fig1}(b) shows the dispersion of SPPs supported by the MWSM plate at the azimuthal angle $\phi=0$, where the nonreciprocal SPPs exhibit an asymmetric dispersion, $\omega(q_x)\neq \omega(-q_x)$. The bulk plasmon dispersion, shown as the gray region, is given by $q^2= \epsilon_{\rm eff} k_0^2$.

The MWSM nanoparticle is modeled as an electric dipole, whose polarizability tensor $\alpha$ is given by the Clausius–Mossotti formula~\cite{HU2023,zhao2012rotational}:
\begin{equation}
 \alpha = 4\pi R^3\frac{\epsilon-1}{\epsilon +2} =
\begin{bmatrix}
    \alpha_{xx} & 0 & \alpha_{xz} \\
    0 & \alpha_{yy} & 0 \\
    -\alpha_{xz} & 0 & \alpha_{zz} \\
  \end{bmatrix} ,
\end{equation}

\begin{figure}[H]
    \centering
    \includegraphics[width=\columnwidth]{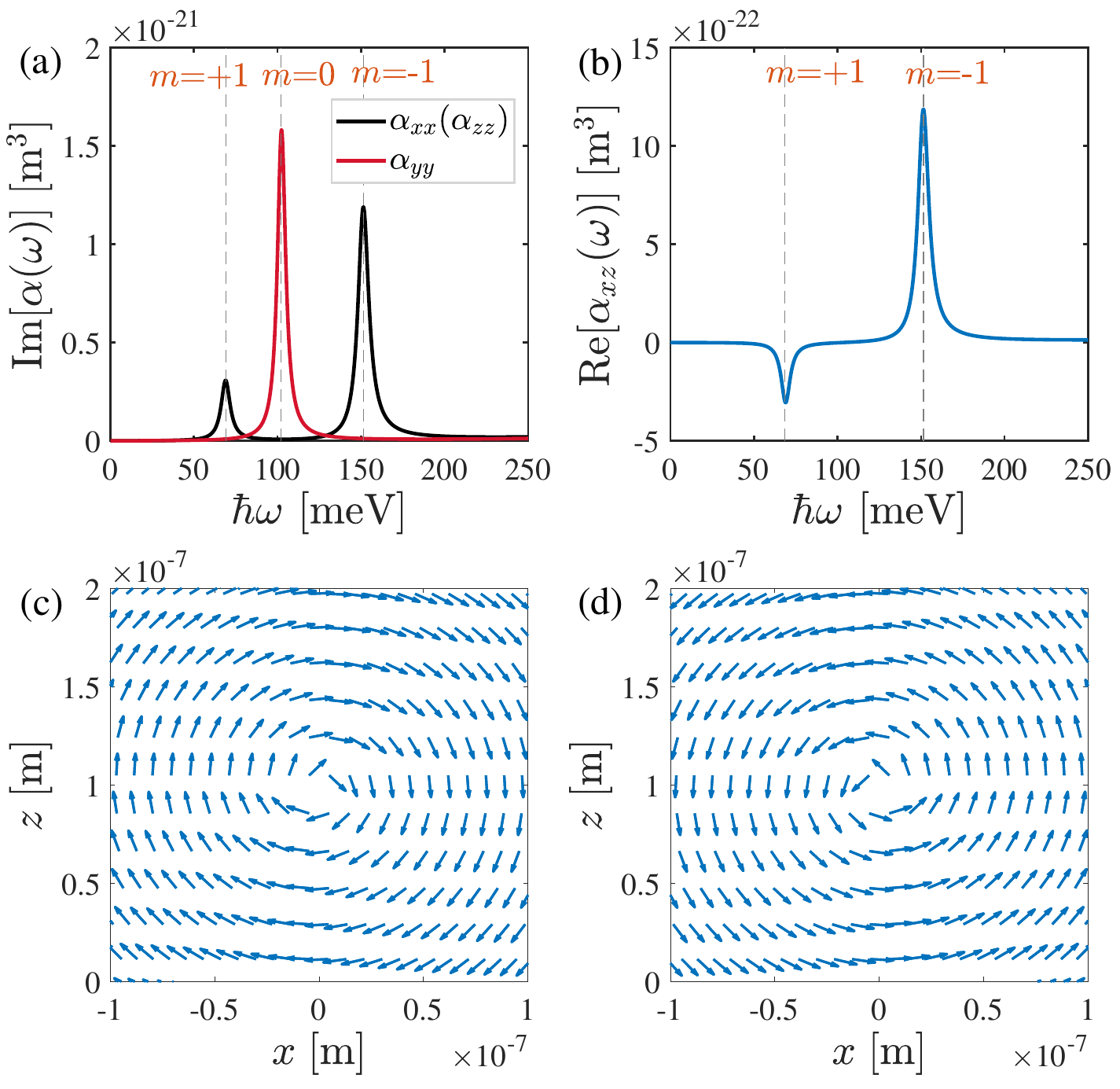} \\
    \caption{(a) Imaginary parts of the diagonal components of the polarizability tensor for the MWSM nanoparticle, showing three distinct dipole resonances at $\omega_{m=0,\pm1}$.
             (b) Real part of the off-diagonal component of the polarizability tensor. Spectral Poynting vector $\mathbf{S_{\omega}}$ (in units of W$\cdot$s/(m$^2$$\cdot$rad)) shown as a normalized vector field in the $x$-$z$ plane in free space at resonance frequencies (c) $\omega_{m=+1}$ and (d) $\omega_{m=-1}$.}
    \label{fig2}
\end{figure}

where $\epsilon$ is given by Eq.~\eqref{Eqe}, and the Weyl node separation of the nanoparticle is fixed along the $+y$ direction. The dipole approximation is valid under the conditions~\cite{messina2018surface} $R \ll d \ll \hbar c/(k_BT)$. Figure~\ref{fig2}(a) displays the imaginary parts of the diagonal components ($\alpha_{xx}$, $\alpha_{yy}$, $\alpha_{zz}$) of the polarizability tensor, revealing three distinct dipole resonances at the frequencies $\omega_{m=0,\pm 1}$~\cite{ott2018circular,Naeimi2025Effi}. Similarly, Fig.~\ref{fig2}(b) shows the real part of the off-diagonal component $\alpha_{xz}$, exhibiting two dipole resonances of opposite sign at frequencies $\omega_{m=\pm1}$. The spectral Poynting vector $\mathbf{S}_{\omega}$ around the MWSM nanoparticle in the absence of the plate can be expressed in a local spherical coordinate system centered at the particle as~\cite{supple}
\begin{align}
\mathbf{S_{\omega}} &= \frac{ \hbar \omega k_{0}^{3} n_p(\omega)}{ 4 \pi^2 r^{2}} \bigg\{\big[{\rm{Im}}(\alpha_{xx}) \left(1+\sin ^{2} \varphi \sin^{2} \theta \right) \notag\\
&+{\rm{Im}}(\alpha_{yy}) \left(\cos ^{2} \theta +\cos^{2} \varphi  \sin^{2} \theta \right) \big] \mathbf{{e}}_{r} - 2 {\rm{Re}}(\alpha_{xz}) \notag\\
&\times \left(\frac{1}{k_{0} r}+\frac{1}{k_{0}^{3} r^{3}}\right) \left(\cos \varphi \, \mathbf{{e}}_{\theta}-\cos \theta \sin \varphi \, \mathbf{{e}}_{\varphi} \right) \bigg\}  \label{S} ,
\end{align}
where $n_p(\omega)=[\exp(\hbar\omega/k_B T_p)-1]^{-1}$ is the Bose-Einstein distribution function, and $(r,\theta,\varphi)$ denote the spherical coordinates with unit basis vectors $\mathbf{e}_{i}$ ($i=r,\theta,\varphi$). Equation~\eqref{S} shows that the off-diagonal polarizability component $\alpha_{xz}$ generates a $\theta$-directed Poynting flux, resulting in a circulating energy flow whose rotation direction (clockwise or counterclockwise) is determined by the sign of ${\rm Re}(\alpha_{xz})$. Figures~\ref{fig2}(c) and \ref{fig2}(d) illustrate the spectral Poynting vectors in the $x$-$z$ plane at frequencies $\omega_{m=+1}$ and $\omega_{m=-1}$, respectively. Specifically, the Poynting vector circulates clockwise at $\omega_{m=+1}$ and counterclockwise at $\omega_{m=-1}$.

\begin{figure}
    \centering
    \includegraphics[width=\columnwidth]{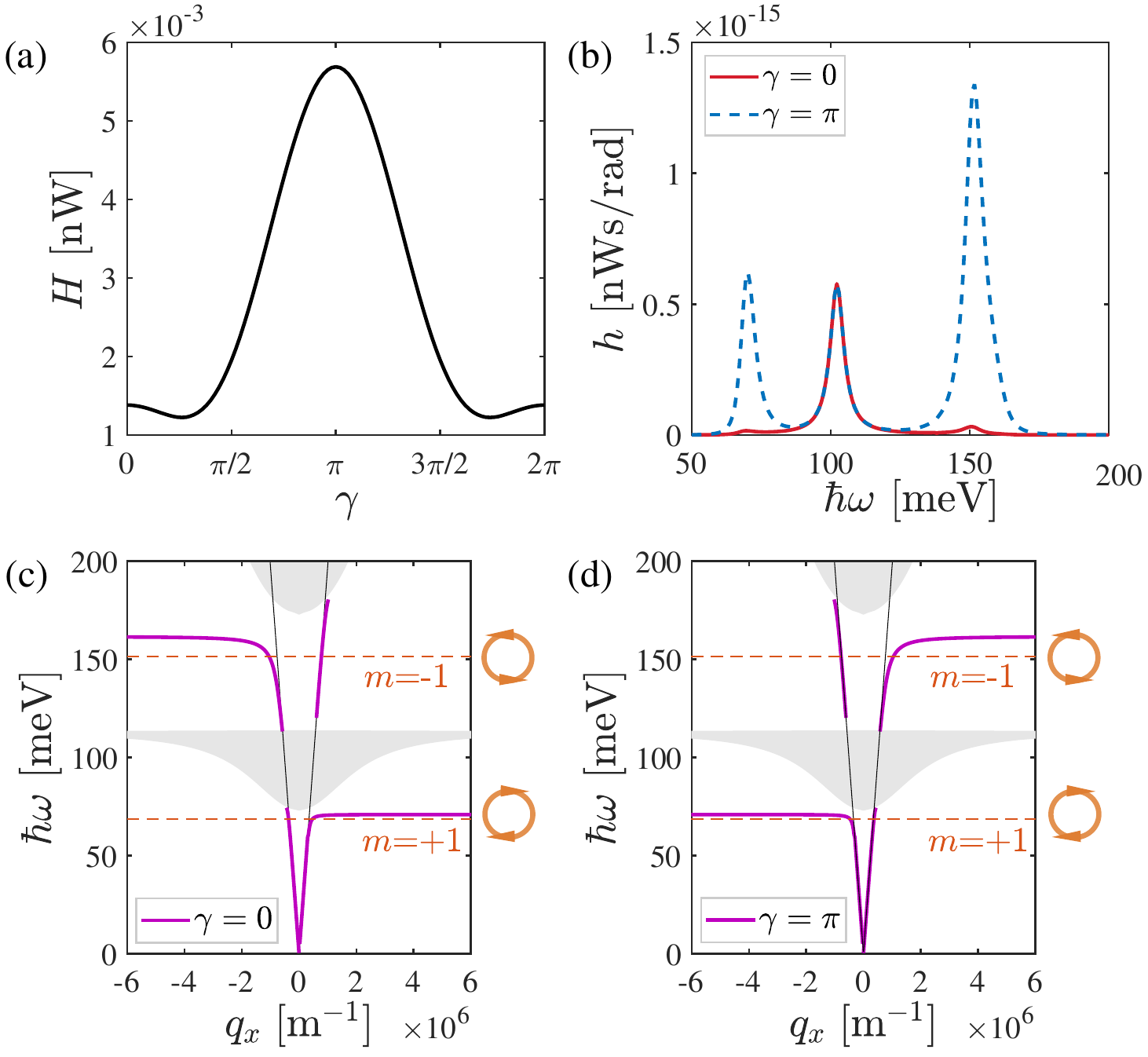} \\
    \caption{(a) Net power transfer as a function of the relative angle $\gamma$.
    (b) Power transfer spectral density $h$  versus  $\hbar\omega$ at $\gamma=0$ and $\gamma=\pi$.
    Schematics illustrating the coupling between the rotational Poynting vector of the particle and the SPPs at $\phi=0$, for (c) $\gamma=0$ and (d) $\gamma=\pi$, respectively. Magenta curves show the dispersion of the nonreciprocal SPPs, while horizontal orange lines mark the particle resonance frequencies $\omega_{m =\pm 1}$; arrows indicate the corresponding Poynting vector rotation directions. }
    \label{fig3}
\end{figure}

\textit{Near-field thermal and momentum transfer.} We first investigate the near-field energy transfer between the  nanoparticle and the plate.
The Weyl node separation of the particle is fixed along the $+y$ direction, while that of the plate is rotated counterclockwise about the $+z$ axis from the same initial direction, introducing a relative angle $\gamma$ between the node separations of the particle and plate as shown in Fig.~\ref{fig1}(a). This configuration enables a controlled study of relative rotation effects arising from the misalignment of the nonreciprocal axes. Figure~\ref{fig3}(a) shows the net power transfer as a function of the relative angle $\gamma$. The results reveal that the power transfer decreases slightly for small $\gamma$, then increases markedly as $\gamma$ approaches $\pi$, displaying a symmetric dependence about $\gamma = \pi$.

Notably, we find that the net power transfer is maximized when the Weyl node separations are antiparallel ($\gamma = \pi$), rather than parallel ($\gamma = 0$). This observation is counterintuitive, as parallel alignment of nonreciprocal axes would typically be expected to enhance mode coupling and promote radiative heat transfer~\cite{peng2021twist}. To elucidate the origin of this phenomenon, we compare the power transfer spectral density for $\gamma = 0$ and $\gamma = \pi$. As shown in Fig.~\ref{fig3}(b), the resonance at $\omega_{m=0}$ remains insensitive to the relative angle, whereas the circular modes at $\omega_{m=\pm1}$ exhibit significant angular dependence. This confirms that the difference in net power transfer between the two configurations is primarily governed by the contributions of the circular modes at $\omega_{m=\pm1}$.

This angular dependence arises from the coupling between the particle's circular modes and the nonreciprocal SPPs. The coupling mechanism is illustrated schematically in Figs.~\ref{fig3}(c) and \ref{fig3}(d), which depict the interplay between the rotation direction of the particle's Poynting vector and the propagation direction of the SPPs for $\gamma = 0$ and $\gamma = \pi$, respectively.
For the parallel case $\gamma = 0 $ [Fig.~\ref{fig3}(c)], the clockwise rotation of the Poynting vector at $\omega_{m=+1}$ is mismatched with the forward ($q_x>0$) SPPs, and the counterclockwise rotation at $\omega_{m=-1}$ is similarly mismatched with the backward ($q_x<0$) SPPs. This double mismatch strongly suppresses the energy flow from the particle to the plate. Conversely, in the antiparallel case $\gamma = \pi$ [Fig.~\ref{fig3}(d)], the rotation directions of the Poynting vectors at $\omega_{m=+1}$ and $\omega_{m=-1}$ align with the backward and forward SPPs, respectively. Therefore, this configuration enables both circular modes to couple strongly to the SPPs simultaneously, leading to the maximum in power transfer. As the relative angle $\gamma$ varies from 0 to $\pi$, the mode-matching condition evolves continuously from the mismatched to the matched configuration, thereby enhancing the net power transfer. This indicates that rotating the MWSM plate provides an efficient means to modulate the near-field energy exchange.

\begin{figure}
    \centering
    \includegraphics[width=\columnwidth]{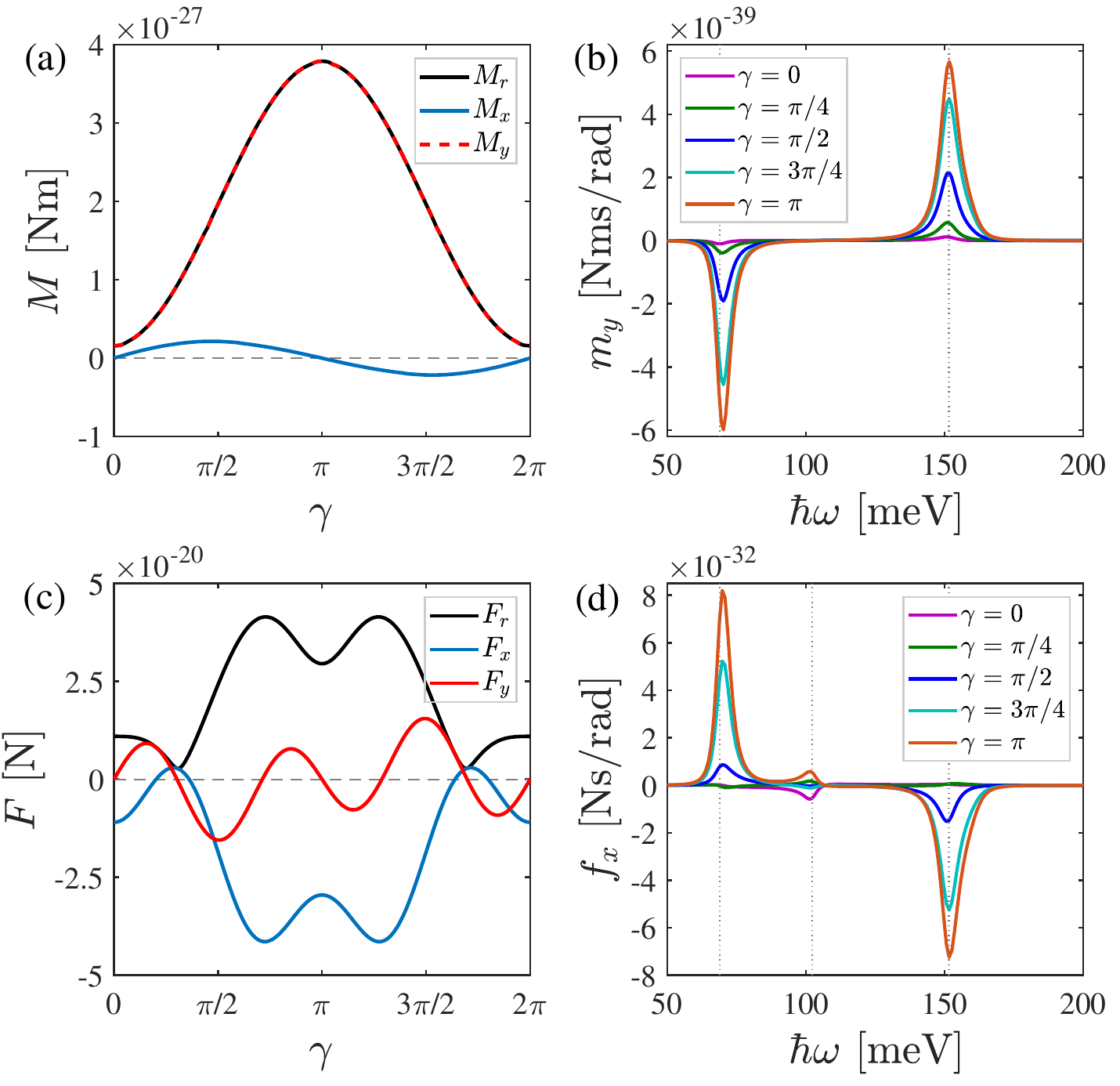} \\
    \caption{(a) Torque components $M_x$, $M_y$, and the resultant magnitude $M_r$ acting on the particle as a function of $\gamma$.
    (b) Torque spectral density $m_y$ versus photon energy $\hbar\omega$ for selected relative angles $\gamma$.
    (c) Force components $F_x$, $F_y$, and the resultant magnitude $F_r$ as a function of $\gamma$.
    (d) Force spectral density $f_x$ versus $\hbar\omega$ for selected relative angles $\gamma$.}
    \label{fig4}
\end{figure}

We next examine the effect of rotation on the angular momentum transfer. Figure~\ref{fig4}(a) shows the torque components $M_x$ and $M_y$, along with the resultant magnitude $M_r=\sqrt{M_x^2+M_y^2}$, as functions of the relative angle $\gamma$. The torque originates from the spin-momentum locking of SPPs propagating in the $x$-$y$ plane, where the transverse spin $\mathbf{s}$ is locked to the momentum $\bm{q}$ via the relation~\cite{VanMechelen2016} $\mathbf{s}\propto\bm{q}\times\hat{z}$. As shown in Fig.~\ref{fig4}(a), the resultant torque $M_r$ is dominated by the component $M_y$, while $M_x$ remains negligible. Moreover, $M_r$ exhibits a unimodal profile symmetric about $\gamma=\pi$, reaching a maximum in the antiparallel configuration. This maximization behavior is consistent with that observed in energy transfer, confirming that the antiparallel alignment is the optimal configuration for enhancing both energy and angular momentum exchange. Figure~\ref{fig4}(b) presents the torque spectral density $m_y$ at various relative angles, revealing that $M_y$ arises from the competing contributions of a negative peak near the dipolar resonance $\hbar\omega_{m=+1}$ and a positive peak near $\hbar\omega_{m=-1}$. As $\gamma$ increases, the coupling between the particle's circular modes and the SPPs gradually strengthens, leading to enhanced angular momentum transfer from the particle to the plate at the resonance frequencies $\omega_{m=\pm1}$. Crucially, although these two modes contribute with opposite signs, the high-frequency mode--with its broader linewidth--ultimately dominates the torque variation. As a result, the resultant torque follows the trend of $M_y$, increasing as $\gamma$ varies from 0 to $\pi$ and reaching an extremum in the antiparallel configuration. 

Figure~\ref{fig4}(c) displays the lateral force components $F_x$, $F_y$, together with the resultant magnitude $F_r=\sqrt{F_x^2+F_y^2}$, as functions of the relative angle $\gamma$. The forces $F_x$ and $F_y$ arise from the asymmetric linear momentum transfer between the particle and the plate along the $x$ and $y$ directions, respectively. Unlike the unimodal behavior of torque, $F_r$ exhibits a nonmonotonic variation with angle $\gamma$. Given that $F_x$ dominates the resultant magnitude at most angles, we analyze the force spectral density $f_x$, as shown in Fig.~\ref{fig4}(d), to elucidate this nonmonotonic behavior. The results show that the $m=-1$ mode, owing to its broader resonant linewidth compared to the $m=+1$ mode, primarily drives the growth of $F_x$ along the $-x$ direction. However, the $m=0$ mode significantly modulates the competition between the $m=\pm 1$ modes. Therefore, the nonmonotonic behavior of $F_r$ results from the competition among the three dipole modes.
These results show that the in-plane torque and force on the MWSM nanoparticle can be continuously tuned by rotating the MWSM plate, providing flexible control over momentum exchange.

\textit{Conclusion.} We have studied near-field radiative transfer of energy and momentum between a magnetic Weyl semimetal nanoparticle and a plate. We have shown that rotating the Weyl node separation of the plate relative to that of the particle provides an efficient means to control radiative energy transfer. This modulation originates from the coupling between the particle-induced rotational Poynting vector and the propagation direction of the nonreciprocal SPPs supported by the plate. In addition, this relative rotation of the Weyl node separations enables tuning of the lateral torque and force acting on the particle.

\begin{acknowledgments}
\textit{Acknowledgements.}
The authors acknowledge the support from the National Natural Science Foundation of China (Grants No. 12574340, No. 12474047, and No. 12374048), the Fund for Shanxi ``1331 Project", and Research Project Supported by Shanxi Scholarship Council of China. This research was partially conducted using the High Performance Computer of Shanxi University.
\end{acknowledgments}

\bigskip
\noindent{$^{*}$gmtang@gscaep.ac.cn}\\
\noindent{$^{\dagger}$zhanglei@sxu.edu.cn}\\
\noindent{$^{\ddag}$chenjun@sxu.edu.cn}

\nocite{Kotov18,wang2018large,Belopolski2019Dis,henkel2002radiation,manjavacas2017lateral,landau2013statistical,Wise22,luo2020radiative}
\bibliography{lateral_fm_WSM}{}

\end{document}


\title{Supplemental Material for ``Modulating near-field radiative energy and momentum
transfer via rotating Weyl semimetals"}
\author{Huimin Zhu,$^{1,4}$ Gaomin Tang,$^{2,*}$ Lei Zhang,$^{1,4,\dagger}$ and Jun Chen$^{3,4,\ddag}$}
\affiliation{$^1$State Key Laboratory of Quantum Optics Technologies and Devices, Institute
  of Laser Spectroscopy, Shanxi University, Taiyuan 030006, China\\
  $^2$Graduate School of China Academy of Engineering Physics, Beijing 100193, China\\
  $^3$State Key Laboratory of Quantum Optics Technologies and Devices, Institute of
  Theoretical Physics, Shanxi University, Taiyuan 030006, China\\
  $^4$Collaborative Innovation Center of Extreme Optics, Shanxi University, Taiyuan
  030006, China}

\bigskip

\begin{abstract}
Section I provides the expression for the optical conductivity of magnetic Weyl semimetals. In Section II, we present the derivation of the spectral densities for radiative power transfer, lateral forces, and lateral torques within the framework of fluctuational electrodynamics. Section III gives the derivation of the spectral Poynting vector around the magnetic Weyl semimetal nanoparticle in free space.
\end{abstract}

\maketitle

\section{I. Optical conductivity of magnetic Weyl semimetals}

The bulk conductivity $\sigma$ of magnetic Weyl semimetals (MWSMs) is obtained from the Kubo-Greenwood formalism using a two-band model with spin degeneracy~\cite{Kotov18}:
\begin{equation}
\begin{aligned}
\sigma = \frac{g r_{s}}{6} \Omega G\left(\frac{\hbar \Omega}{2}\right)+i \frac{g r_{s}}{6 \pi} \Bigg\{ \frac{4}{\hbar^{2} \Omega}\left[E_F^{2}+\frac{\pi^{2}}{3}\left(k_{\mathrm{B}} T\right)^{2}\right]
 +8 \Omega \int_{0}^{E_c} \frac{G(E)-G(\hbar \Omega / 2)}{(\hbar \Omega)^{2}-4 E^{2}} E ~dE \Bigg\},
\end{aligned}
\end{equation}
where $r_s = e^2/(4\pi\epsilon_0\hbar v_F)$ is the effective fine-structure constant with Fermi velocity $v_F$, $g$ is the number of Weyl points, $\Omega = \omega + i 2\pi \tau^{-1}$ with the Drude damping rate $\tau^{-1}$, $G(E) = n(-E) - n(E)$ with the Fermi distribution function $n(E)$, $E_F$ is the chemical potential, and $E_c$ is the cutoff energy.
The parameters are as follows~\cite{Dong2021}: $\epsilon_b = 6.2$, $g = 2$, $\tau = 1000\,$fs, $E_F = 0.15\,$eV, $E_c = 3 E_F$, $b = 10^9\,$m$^{-1}$, and $v_F = 2 \times 10^5\,$m/s. The parameters are close to the reported values for $\mathrm{Co_3Sn_2S_2}$~\cite{wang2018large} and $\mathrm{Co_2MnGa}$~\cite{Belopolski2019Dis}.

\section{II. Derivation of the spectral densities of power transfer, forces and torques}

Within the framework of fluctuational electrodynamics, the total power from the particle to the plate $H$, the lateral forces $F_j$ with $j=x,y$ and the lateral torques $M_j$ can be expressed as
\begin{equation}
 P =
 \int_{0}^{\infty} \frac{d\omega}{2\pi} p(\omega) =
 \frac{1}{2}\int_{-\infty}^{\infty} \frac{d\omega}{2\pi} p(\omega),
\end{equation}
with $P\in \{H, F_x,F_y, M_x,M_y\}$ and the corresponding spectral densities $p\in \{h,f_x,f_y,m_x,m_y\}$.
The spectral densities can be expressed as~\cite{henkel2002radiation,manjavacas2017lateral}
\begin{align}
  h(\omega)&= 2\int_{-\infty}^{\infty} \frac{d\omega'}{2\pi}\omega {\rm Im}\Big[\Big\langle p_{j}^{{\text{fl}}^*}(\omega) E_{j}^{\text{ind}}(\omega')+ p_{j}^{{\text{ind}}^*}(\omega) E_{j}^{{\text{fl}}}(\omega')\Big\rangle\Big]
  e^{i(\omega-\omega')t},
\label{Eqh} \\
  f_j(\omega)&= 2\int_{-\infty}^{\infty} \frac{d\omega'}{2\pi} {\rm Re}\Big[\Big\langle p_k^{\text{fl}}(\omega) \partial_j E_k^{\text{ind}^*}(\omega')+ p_k^{\text{ind}}(\omega) \partial_j E_k^{{\text{fl}}^*}(\omega')\Big\rangle\Big] e^{i(\omega'-\omega)t},
\label{Eqf} \\
  m_j(\omega) &= 2\int_{-\infty}^{\infty} \frac{d\omega'}{2\pi} \epsilon_{jkl} {\rm Re}\Big[\Big\langle p_k^{{\text{fl}}}(\omega) E_l^{{\text{ind}}^*}(\omega')+ p_k^{\text{ind}}(\omega) E_l^{{\text{fl}}^*}(\omega')\Big\rangle\Big] e^{i(\omega'-\omega)t},
\label{Eqm}
\end{align}
where the Einstein summation convention is applied with $j ,k, l\in \{x, y, z\}$, and $\langle\cdots\rangle$ denotes the statistical ensemble average. $E^{\mathrm{fl}}$ is the electric field generated by current fluctuations on the surface, $p^{\mathrm{fl}}$ is the fluctuating dipole moment,
$E^{\mathrm{ind}}$ is the field induced by the particle dipole moment fluctuations, and $p^{\mathrm{ind}}$ is the dipole moment induced in the particle by the environment field fluctuations. We assume that the nanoparticle is located at $\mathbf{r}$, and the dependence of $\mathbf{r}$ for the quantities on the right-hand sides are not written out explicitly.
The dipole moment induced by the electric-field fluctuations is given by
\begin{equation} \label{pind}
p_j^{\text{ind}}(\mathbf{r}) = \epsilon_0 \alpha_{j k} E_k^{\mathrm{fl}}(\mathbf{r}) ,
\end{equation}
where $\alpha$ is the polarizability tensor of the particle.
The induced field by the dipole fluctuations can be expressed as
\begin{equation} \label{Eind}
E_{j}^{\mathrm{ind}}(\mathbf{r}_2, \omega)=\omega^2 \mu_{0} G_{j k}(\mathbf{r}_2, \mathbf{r}, \omega) p_{k}^{\mathrm{fl}}(\mathbf{r}, \omega).
\end{equation}
The antisymmetrized correlations functions of the fluctuating quantities satisfy the fluctuation-dissipation relations with~\cite{landau2013statistical}
\begin{align}
\Big\langle p_j^{\mathrm{fl}}(\mathbf{r},\omega) p_k^{\mathrm{fl}^*}(\mathbf{r}, \omega') \Big\rangle = \frac{\hbar \epsilon_0}{2i} \left(\alpha_{j k}-\alpha_{k j}^* \right)
 \bigg[n_p(\omega)+\frac{1}{2} \bigg] 4 \pi\delta(\omega-\omega'),
\end{align}
and
\begin{align}
\Big\langle E_j^{\mathrm{fl}}(\mathbf{r}_1, \omega)E_k^{\mathrm{fl}^{*}}(\mathbf{r}_2, \omega')\Big\rangle = \frac{1}{2i}\big[ G_{j k}(\mathbf{r}_1,\mathbf{r}_2, \omega)- G_{k j}^*(\mathbf{r}_2, \mathbf{r}_1, \omega) \big]   \big(\hbar \mu_0 \omega^2 \big)\bigg[n_e( \omega )+\frac{1}{2} \bigg]
4 \pi \delta(\omega-\omega')\delta(\mathbf{r}_2-\mathbf{r}_1),
\end{align}
where $G_{j k}(\mathbf{r}_1,\mathbf{r}_2, \omega)$ is the Green's function in the dipole-plate geometry given by Eqs.~(8)-(10) of the main text, and the Bose-Einstein distribution functions are given by
\begin{equation}
  n_{p/e}(\omega) = \frac{1}{\exp\left( \frac{\hbar\omega}{k_B T_{p/e}} \right)-1} .
\end{equation}
The radiation power spectral density can be obtained as
\begin{align}
h(\omega) = & 2\int_{-\infty}^{\infty} \frac{d\omega'}{2\pi}\omega {\rm Im}\Big[\Big\langle p_{j}^{{\text{fl}}^*}(\omega) E_{j}^{\text{ind}}(\omega')+ p_{j}^{{\text{ind}}^*}(\omega) E_{j}^{{\text{fl}}}(\omega')\Big\rangle\Big]
e^{i(\omega-\omega')t}  \notag \\
= & 2 \omega {\rm Im}\Big[\omega^2 \mu_{0}\Big\langle p_j^{\text{fl}^*}(\mathbf{r}, \omega) p_k^{\text{fl}}(\mathbf{r}, \omega) \Big\rangle G_{j k}(\mathbf{r},\mathbf{r},\omega)+\epsilon_0 \alpha_{jk}^*\Big\langle E_k^{\text{fl}^*}(\mathbf{r}, \omega) E_j^{\text{fl}}(\mathbf{r}, \omega)\Big\rangle\Big]  \notag \\
=& 4 \hbar\omega k_0^2 \delta n(\omega) \int \frac{d^2 {\bm q}}{(2\pi)^2} \big[ {\rm Im}(\alpha_{xx}) {\rm Im} \left( G_{xx} +G_{zz} \right)+{\rm Im}(\alpha_{yy}) {\rm Im} \left(G_{yy} \right)+ 2{\rm Re}(\alpha_{xz}) {\rm Re}\left(G_{xz} \right) \big],
\end{align}
we abbreviate the Green's function $G_{ij}({\bm q}, d, d, \omega)$ as $G_{ij}$ where $i$,$j$ $\in$\{$x$, $y$, $z$\}. The in-plane wavevector is ${\bm q}=(q_x, q_y)$ with $q = |\bm{q}|$, and the photon-occupation difference is
\begin{equation}
  \delta n(\omega) = n_p(\omega) - n_e(\omega).
\end{equation}
The spectral density of force along the $x$ direction, $f_x$, can be obtained as
\begin{align}\label{fx}
 f_x(\omega)&= 2\int_{-\infty}^{\infty} \frac{d\omega'}{2\pi} {\rm Re}\Big[ \Big\langle p_k^{\text{fl}}(\omega) \partial_x E_k^{\text{ind}^*}(\omega')+ p_k^{\text{ind}}(\omega) \partial_x E_k^{{\text{fl}}^*}(\omega')\Big\rangle \Big] e^{i(\omega'-\omega)t} \notag \\
 =& 2 {\rm Re}\Big[\Big\langle p_k^{\text{fl}}(\mathbf{r},\omega) \partial_{x}\big(\omega^2 \mu_0 G_{k j}(\mathbf{r},\mathbf{r},\omega) p_j^{\text{fl}}(\mathbf{r},\omega)\big)^* +\epsilon_0 \alpha_{kj} E_j^{\text{fl}}\left(\mathbf{r}, \omega\right) \partial_x E_k^{\text{fl}^*}(\mathbf{r}, \omega)\Big\rangle \Big]  \notag \\
 =&-4\hbar k_0^2 \delta n(\omega) \int \frac{d^2 {\bm q}}{(2\pi)^2} \Big\{q_{x} \big[ {\rm Im}(\alpha_{xx}){\rm Im} \left( G_{xx}+G_{zz} \right) +{\rm Im}(\alpha_{yy}) {\rm Im}\left(G_{yy} \right)+ 2{\rm Re}\left(\alpha_{xz} \right) {\rm Re} \left( G_{xz} \right) \big] \Big\} .
\end{align}
In above, $\partial_x$ represents partial derivative in the $x$ direction of position $\mathbf{r}$. For the spectral density of force along the $y$ direction, $f_y$, just replace $q_{x}$ in Eq.~\eqref{fx} with $q_{y}$.

The spectral density of torque along the $x$ direction, $m_x$, can be obtained as
\begin{align}
 m_x(\omega) = & 2\int_{-\infty}^{\infty} \frac{d\omega'}{2\pi} \epsilon_{xkl} {\rm Re}\Big[\Big\langle p_k^{{\text{fl}}}(\omega) E_l^{{\text{ind}}^*}(\omega')+ p_k^{\text{ind}}(\omega) E_l^{{\text{fl}}^*}(\omega')\Big\rangle\Big] e^{i(\omega'-\omega)t}  \notag \\
 =&2\omega^2 \mu_{0} {\rm Re}\Big[\Big\langle p_y^{\text{fl}}(\mathbf{r},\omega) G_{z j}^*(\mathbf{r},\mathbf{r},\omega) p_j^{\text{fl}^*}(\mathbf{r},\omega)  \Big\rangle  -\Big\langle p_z^{\text{fl}}(\mathbf{r}, \omega) G_{y k}^*(\mathbf{r},\mathbf{r},\omega) p_k^{\text{fl}^*}(\mathbf{r},\omega) \Big\rangle \Big] \notag \\
 & + 2\epsilon_0 {\rm Re}\Big[\alpha_{yj}(\omega) \Big\langle E_j^{\text{fl}}(\mathbf{r},\omega) E_z^{\text{fl}^*}(\mathbf{r},\omega) \Big\rangle
 -\alpha_{zk}(\omega)\Big\langle E_k^{\text{fl}}(\mathbf{r},\omega) E_y^{\text{fl}^*}(\mathbf{r},\omega) \Big\rangle\Big] \notag \\
 =&4\hbar k_0^2 \delta n(\omega) \int \frac{d^2 {\bm q}}{(2\pi)^2} \big[ {\rm Im}(\alpha_{xx}+\alpha_{yy}){\rm Re}\left(G_{zy}\right) -{\rm Re}(\alpha_{xz}) {\rm Im}\left( G_{yx} \right) \big].
\end{align}
Similarly, the torque spectral density $m_y$ along the $y$ direction can be obtained as
\begin{align}
 m_y\omega) = & 2\int_{-\infty}^{\infty} \frac{d\omega'}{2\pi} \epsilon_{ykl} {\rm Re}\Big[\Big\langle p_k^{{\text{fl}}}(\omega) E_l^{{\text{ind}}^*}(\omega')+ p_k^{\text{ind}}(\omega) E_l^{{\text{fl}}^*}(\omega')\Big\rangle\Big] e^{i(\omega'-\omega)t}  \notag \\
=& 2\omega^2 \mu_{0} {\rm Re}\Big[\Big\langle p_z^{\text{fl}}(\mathbf{r},\omega) G_{x j}^*(\mathbf{r},\mathbf{r},\omega) p_j^{\text{fl}^*}(\mathbf{r},\omega)  \Big\rangle  -\Big\langle p_x^{\text{fl}}(\mathbf{r}, \omega) G_{z k}^*(\mathbf{r},\mathbf{r},\omega) p_k^{\text{fl}^*}(\mathbf{r},\omega) \Big\rangle \Big] \notag \\
& +2\epsilon_0 {\rm Re}\Big[\alpha_{zj} \Big\langle E_j^{\text{fl}}(\mathbf{r},\omega) E_x^{\text{fl}^*}(\mathbf{r},\omega) \Big\rangle
 -\alpha_{xk}\Big\langle E_k^{\text{fl}}(\mathbf{r},\omega) E_z^{\text{fl}^*}(\mathbf{r},\omega) \Big\rangle\Big]  \notag \\
 =&4\hbar k_0^2 \delta n(\omega) \int \frac{d^2 {\bm q}}{(2\pi)^2}  \big[2 {\rm Im}(\alpha_{xx})  {\rm Re}\left(G_{xz}\right)  + {\rm Re}(\alpha_{xz}) {\rm Im}\left(G_{xx} +G_{zz} \right) \big].
\end{align}

\section{III. Derivation of the spectral Poynting vector around the MWSM nanoparticle}
Following the model developed in work~\cite{ott2018circular}, we calculate Poynting vector $\mathbf{S}$ around the MWSM nanoparticle in free-space without the MWSM plate. The mean Poynting vector is calculated as
\begin{align}
\mathbf{S} =
\big\langle \mathbf{E} \left(\mathbf{r}, t\right) \times \mathbf{H}\left(\mathbf{r}, t\right) \big\rangle
= \int_{-\infty}^{\infty} \frac{d \omega}{2 \pi}
\int_{-\infty}^{\infty} \frac{d \omega'}{2 \pi}
\big\langle \mathbf{E} \left(\mathbf{r}, \omega\right) \times \mathbf{H}^{*}\left(\mathbf{r}, \omega'\right)
 e^{-i(\omega - \omega')t} \big\rangle ,
\end{align}
where $\langle \cdots \rangle$ denotes the symmetrized correlation and we have used the relation $
\mathbf{H}^{*}(\mathbf{r}, \omega)
=\mathbf{H}(\mathbf{r}, -\omega)$.
By considering the spectral Poynting vector with only positive frequency $\mathbf{S_{\omega}}$ with
\begin{equation}
\mathbf{S} = \int_{0}^{\infty} \frac{d \omega}{2\pi} \mathbf{S_{\omega}},
\end{equation}
we have
\begin{equation}
\mathbf{S_{\omega}} = 2\int_{-\infty}^{\infty} \frac{d \omega'}{2 \pi}
 \big\langle \mathbf{E} \left(\mathbf{r}, \omega\right) \times \mathbf{H}^{*}\left(\mathbf{r}, \omega'\right)
 e^{i(\omega' - \omega)t} \big\rangle  .
\end{equation}
The induced electric and magnetic fields by the dipole fluctuations can be expressed as
\begin{equation}
\mathbf{E}(\mathbf{r}, \omega) =\omega^2 \mu_0 G^E (\mathbf{r}, \mathbf{r}', \omega) \mathbf {p}^{\rm fl}(\mathbf{r}', \omega), \quad
\mathbf{H}(\mathbf{r}, \omega) =\omega^2 \mu_0 G^H(\mathbf{r}, \mathbf{r}', \omega) \mathbf{p}^{\rm fl}(\mathbf{r}', \omega),
\end{equation}
where the electric and the magnetic Green's functions are defined as~\cite{ott2021thermal,luo2020radiative,messina2018surface}
\begin{align}
G^E \left(\mathbf{r}, \mathbf{r}', \omega\right) &= \frac{e^{i k_{0} r}}{4 \pi r} \left[ \left(1+\frac{i k_{0} r-1}{k_{0}^{2} r^{2}}\right) \mathbf{I} + \left(\frac{3-3 i k_{0} r-k_{0}^{2} r^{2}}{k_{0}^{2} r^{2}}\right) \mathbf{e}_{r} \otimes \mathbf{e}_{r} \right], \\
G^H \left(\mathbf{r}, \mathbf{r}', \omega\right) &= \sqrt{\frac{\epsilon_{0}}{\mu_{0}}} \frac{e^{i k_{0} r}}{4 \pi r} \left(1+\frac{i}{k_{0} r}\right) \left(\mathbf{e}_{\varphi} \otimes \mathbf{e}_{\theta} - \mathbf{e}_{\theta} \otimes \mathbf{e}_{\varphi} \right) .
\end{align}
Here, the particle is located at $\mathbf{r}' = (0,0,d)$, where $d$ is the distance between the particle and the plate, and the observation point is at $\mathbf{r} = (x,y,z)$. We introduce a local spherical coordinate system $(r, \theta, \varphi)$ centered at the particle position $\mathbf{r}'$. The relative position vector is given by $\mathbf{r} - \mathbf{r}' = (x, y, z-d)$, and the relative distance is denoted by $r = |\mathbf{r} - \mathbf{r}'| = \sqrt{x^2 + y^2 + (z-d)^2}$. The angle $\theta$ represents the polar angle with respect to the $z$ axis, and $\varphi$ is the azimuthal angle in the $x$-$y$ plane:
\begin{align}
\theta = \arccos\left(\frac{z-d}{r}\right),\quad
\varphi = {\arctan}(y/x).
\end{align}
The corresponding unit vectors are given by
\begin{align}
\mathbf{e}_r = (\sin\theta\cos\varphi,\ \sin\theta\sin\varphi,\ \cos\theta), \quad
\mathbf{e}_\theta = (\cos\theta\cos\varphi,\ \cos\theta\sin\varphi,\ -\sin\theta), \quad
\mathbf{e}_\varphi = (-\sin\varphi,\ \cos\varphi,\ 0).
\end{align}
With these expressions, the spectral Poynting vector is calculated as follows
\begin{align}
{S_{j}}(\omega) =4\hbar\omega^2 k_0^2 \mu_0 \epsilon_{j k l} {\rm Re} \Bigg\{  n_p(\omega) G^E\left(\mathbf{r}, \mathbf{r}', \omega\right) \frac{{\alpha} -{\alpha}^\dagger}{2i} \Big[G^H\left(\mathbf{r}, \mathbf{r}', \omega\right)\Big]^\dagger \Bigg\}_{k l}  ,
\end{align}
where $\epsilon_{j k l}$ is the Einstein summation convention with $j,k,l \in\{x, y, z\}$. The spectral Poynting vector is expressed in local spherical coordinates as
\begin{align}
\mathbf{S}_\omega=&\frac{ \hbar \omega k_{0}^{3} n_p(\omega)}{  4 \pi^2 r^{2}} \bigg\{\big[{\rm{Im}}(\alpha_{xx}) \left(1+\sin ^{2} \varphi \sin ^{2} \theta \right)+{\rm{Im}}(\alpha_{yy}) \left(\cos ^{2} \theta +\cos ^{2} \varphi  \sin ^{2} \theta \right) \big] \mathbf{{e}}_{r}  \notag \\
&  -2 {\rm{Re}}(\alpha_{xz}) \left(\frac{1}{k_{0} r}+\frac{1}{k_{0}^{3} r^{3}}\right)  \left(\cos \varphi \, \mathbf{{e}}_{\theta}-\cos \theta \sin \varphi \, \mathbf{{e}}_{\varphi} \right) \bigg\} ,
\end{align}
where $n(\omega)= \left[\exp(\hbar\omega/{k_B T_p})-1 \right]^{-1}$ is the Bose-Einstein distribution function at temperature $T_p$.

\bibliography{Supplemental_Material}{}